\documentclass[usenatbib]{mn2e}
\usepackage{amssymb}
\usepackage{color}
\usepackage{graphicx}

\newcommand{\be}{\begin{equation}}
\newcommand{\ee}{\end{equation}}
\newcommand{\ba}{\begin{eqnarray}}
\newcommand{\ea}{\end{eqnarray}}
\newcommand{\lp}{\left(}
\newcommand{\rp}{\right)}

\newcommand{\Pdot}{\dot{P}}
\newcommand{\Edot}{\dot{E}}
\newcommand{\Lx}{L_{\mathrm{x}}}
\newcommand{\tauc}{\tau_{\mathrm{c}}}

\title[Radio unification of magnetars and pulsars]{More than meets the eye:
magnetars in disguise}
\author[W.C.G. Ho]{Wynn C. G. Ho,$^1$\thanks{Email: wynnho@slac.stanford.edu}
\\
$^1$School of Mathematics, University of Southampton, Southampton, SO17 1BJ
}
\date{Accepted 2012 October 30. Received 2012 October 4;
 in original form 2012 August 1}

\begin{document}
\pagerange{\pageref{firstpage}--\pageref{lastpage}} \pubyear{2012}

\maketitle

\label{firstpage}

\begin{abstract}
It has recently been proposed that radio emission from magnetars can be
evaluated using a ``fundamental plane'' in parameter space between
pulsar voltage gap and ratio of X-ray luminosity $\Lx$ to rotational
energy loss rate $\Edot$.
In particular, radio emission from magnetars will occur if $\Lx/\Edot<1$
and the voltage gap is large, and there is no radio emission if $\Lx/\Edot>1$.
Here we clarify several issues regarding this fundamental plane, including
demonstrating that the fundamental plane is not uniquely defined.
We also show that, if magnetars and all other pulsars are different
manifestations of a unified picture of neutron stars, then pulsar radio
activity (inactivity) appears to be determined by the ratio
$\Lx/\Edot\lesssim1$ ($\Lx/\Edot\gtrsim 1$), although observational bias
and uncertainty in the ratio for some sources may still invalidate this
conclusion.
Finally, we comment on the use of other pulsar parameters that are
constructed from the three observables:
spin period $P$, period derivative $\Pdot$, and $\Lx$.
\end{abstract}


\maketitle

\section{Introduction} \label{sec:intro}

Anomalous X-ray pulsars (AXPs) and soft gamma-ray repeaters (SGRs) form the
magnetar class of neutron stars, i.e., neutron stars which possess superstrong
magnetic fields ($B\gtrsim 10^{14}\mbox{ G}$) in most cases.
Their strong fields likely power the activity seen in these objects
(see \citealt{woodsthompson06,mereghetti08}, for review;
see McGill SGR/AXP Online
Catalog\footnote{http://www.physics.mcgill.ca/$\sim$pulsar/magnetar/main.html}
for observational details).
Two notable (and formerly defining) properties of magnetars are
their high (as compared to that of other neutron stars of a similar age)
observed X-ray luminosities $\Lx$ in quiescence and
their non-detection at radio wavelengths.
The first suggests that heat generated from the decay of a strong magnetic
field is the source of their bright X-ray luminosity
\citep{thompsonduncan96,heylkulkarni98,colpietal00,aguileraetal08}
since $\Lx$ is greater than that available from their rotational and
thermal reservoirs.
The second suggests that magnetars do not emit in radio.
However, recent observations have brought these characteristics into
question, in particular, the discovery of X-ray luminosities lower than
spin-down luminosities (or rate of rotational energy loss $\Edot$) for,
and radio emission from, several magnetars
(see \citealt{reaetal12}, and references therein).
The blurring of distinctions between magnetars and normal rotation-powered
pulsars suggests magnetars may simply be a different manifestation of normal
pulsars (see, e.g., \citealt{kaspi10,pernapons11,ponsperna11}),
although radio emission from magnetars do show some behavior that are
different than radio emission from normal pulsars (see discussion in
Sec.~2 of \citealt{reaetal12}, and references therein).

In light of these latest discoveries, \citet{reaetal12} examine the
observed properties of radio active and inactive magnetars.
They notice that all radio magnetars have X-ray efficiency $\Lx/\Edot<1$.
They also calculate the electric potential difference across the magnetic
pole $\Delta\Phi$ (or voltage gap) for magnetars and normal pulsars
and find an apparent anti-correlation between voltage gap and X-ray efficiency
for magnetars.
They then conduct simulations which can produce an anti-correlation between
$\Delta\Phi$ and $\Lx/\Edot$, depending on the neutron star magnetic field
at birth.
They conclude that there exists a fundamental plane
($\Delta\Phi$ versus $\Lx/\Edot$) for radio magnetars,
in which a magnetar will be radio active if $\Lx/\Edot<1$ and the voltage
gap is large and radio inactive if $\Lx/\Edot>1$.

Here we point out that one needs to be careful about claiming trends and
correlations between parameters (e.g., $\Delta\Phi$ and $\Lx/\Edot$)
when, in fact, these parameters are not entirely independent.
We also extend the analysis of \citet{reaetal12} by considering magnetars
and other X-ray bright pulsars within the unified picture of neutron stars,
as outlined in \citet{kaspi10}.

In Section~\ref{sec:model}, we briefly describe the standard model for
pulsars and some of the basic equations derived from this model,
as well as mention relevant recent works.
In Section~\ref{sec:results}, we show results from using these
equations and the observed properties of magnetars and other pulsars.
We summarize our results and discuss their implications in
Section~\ref{sec:discuss}.

\section{Basic equations of pulsar standard model} \label{sec:model}

For rotation-powered pulsars,
the conventional picture is that a pulsar emits radiation at a cost to its
rotational energy,
and the rate at which its rotational energy decreases is given
by\footnote{There is a typo in the numerical value for rotational energy
loss given in \citet{reaetal12}.}
\be
\Edot = 4\pi^2I\frac{\Pdot}{P^3} = \frac{2\pi^2I}{P^2\tauc}
 = 4.0\times 10^{46}\mbox{ ergs s$^{-1}$}I_{45}\frac{\Pdot}{P^3},
 \label{eq:edot}
\ee
where $I$ is the neutron star moment of inertia,
$P$ and $\Pdot$ are the spin period and spin period derivative, respectively,
and $I_{45}=I/10^{45}\mbox{ g cm$^2$}$.
The characteristic age of the pulsar (often used as an estimate of true age)
is defined to be
\be
\tauc = \frac{P}{2\Pdot}. \label{eq:tauc}
\ee
Note that equations~(\ref{eq:edot}) and (\ref{eq:tauc}) do not depend on a
specific emission mechanism.

By assuming that the rotational energy is lost through magnetic dipole
radiation, the pulsar magnetic field $B$ can be inferred from
observables $P$ and $\Pdot$, i.e.,
\be
P\Pdot = (\gamma/2) B^2 \quad\Leftrightarrow\quad
 B \sim 6.4\times 10^{19}\mbox{ G }(P\Pdot)^{1/2}, \label{eq:magb}
\ee
where $\gamma=4\pi^2R^6\sin^2\alpha/3c^3I
 = 4.884\times 10^{-40}\mbox{ s G$^{-2}$ }R_6^6I_{45}^{-1}\sin^2\alpha$,
$R$ is the neutron star radius, $\alpha$ is the angle between the stellar
rotation and magnetic axes, and $R_6=R/10^6\mbox{ cm}$ \citep{gunnostriker69}.
For radio emission, what is likely to be important is the electric potential
difference across the magnetic pole, the maximum of which is approximately
given by\footnote{Eq.~(2) of \citet{reaetal12} gives the potential
difference in statvolts; however the numerical value is actually in volts;
similarly, the $y$-axes in both panels of Fig.~2 of \citet{reaetal12} are
in volts, not statvolts.\label{foot:volt}}
\citep{goldreichjulian69}
\ba
\Delta\Phi &=& \frac{BR}{2}\lp\frac{2\pi R}{cP}\rp^2
 = \lp\frac{3\Edot}{2c\sin^2\alpha}\rp^{1/2}
 \sim (3\Edot/2c)^{1/2}
 \nonumber\\
 &=& 1.4\times 10^{18}\mbox{statvolts } (\Pdot/P^3)^{1/2}. \label{eq:potent}
\ea
Clearly this voltage gap is trivially related to the rotational energy
loss, i.e., $\Delta\Phi^2\propto\Edot$.  Therefore measuring $\Edot$ is
equivalent to determining $\Delta\Phi$ in the standard model for pulsar
radio emission.
In the following, we will only consider $\Edot$ since it is a more
fundamental parameter
(i.e., it is derived from only $I$ and two observables $P$ and $\Pdot$).

We note that the standard model described above follows from the early works
of, e.g., \citet{goldreichjulian69,rudermansutherland75,aronsscharlemann79}.
More recently, the numerical calculations of pulsar magnetospheres
by \citet{spitkovsky06} find a modified equation for the energy emitted
by an inclined, rotating neutron star.
This leads to an inferred magnetic field at the magnetic equator
[c.f. eq.~(\ref{eq:magb}), at the magnetic pole]
\be
B\approx 2.6\times 10^{19}\mbox{ G }[P\Pdot/(1+\sin^2\alpha)]^{1/2}
 \label{eq:magbspit}
\ee
and a voltage gap $\Delta\Phi\approx\{\Edot/[c(1+\sin^2\alpha)]\}^{1/2}$
[c.f. eq.~(\ref{eq:potent})].
Furthermore, recent theoretical studies of the magnetosphere indicate that
magnetar radio activity originates from closed magnetic field lines, in
contrast to normal pulsar radio activity, and this difference could contribute
to their differing observed properties; they also find the actual voltage gap
to be much lower than that given by eq.~(\ref{eq:potent}), in particular
$\Delta\Phi\sim 10^3m_{\mathrm{e}}c^2\sim 10^6\mbox{ statvolts}$ (see, e.g.,
\citealt{beloborodovthompson07,beloborodov12};
see also discussion in Sec.~4 of \citealt{reaetal12}, and references therein).

\section{Comparison to observations} \label{sec:results}

We gather pulsar periods, period derivatives, spin-down luminosities,
characteristic ages, and X-ray luminosities from the following:
For rotation-powered radio pulsars,
values for $\Edot$, $\tauc$, and $\Lx$ are taken from \citet{becker09}.
For magnetars, values for $\Edot$ and $\tauc$ are taken from the
McGill SGR/AXP Online Catalog.
For all other sources, values for $\Edot$ and $\tauc$ are calculated
from $P$ and $\Pdot$, which are obtained from the ATNF Pulsar Catalogue
\citep{manchesteretal05}\footnote{http://www.atnf.csiro.au/research/pulsar/psrcat/},
except in the case of 1E~1207.4$-$5209, where we consider two values of $\Edot$
since there are two measured values of $\Pdot$ \citep{halperngotthelf11}.
For magnetars, values for $\Lx$ are taken from
\citet{durantvankerkwijk06,gelfandgaensler07,munoetal07,reaetal07,reaetal09b,reaetal12b,espositoetal08,bernardinietal09,halperngotthelf10,andersonetal12},
as well as \citet{kaminkeretal09,ngkaspi11}, and references therein.
For high magnetic field pulsars
(three of which are rotating radio transients; \citealt{mclaughlinetal06}),
$\Lx$ upper limits are taken from \citet{ngkaspi11},
while $\Lx$ is calculated from those with measured temperature $T$ and radius
$R$ (see \citealt{zhuetal11}, and references therein).
For the ROSAT isolated neutron stars,
$\Lx$ is calculated using $T$ and $R$ that are taken from
\citet{kaplanvankerkwijk11} and \citet{zhuetal11}, and references therein.
For PSR~J0726$-$2612, $\Lx$ is calculated using $T$ and $R$ that taken from
\citet{speagleetal11}.
For the central compact objects, $\Lx$ upper limits (for emission from the
entire stellar surface) are taken from
\cite{delucaetal04,gotthelfetal10,halperngotthelf10a}.
Note that the X-ray luminosities are used for illustrative purposes only
since they are not calculated consistently between sources,
and some sources are variable and have large distance
uncertainties\footnote{For example, PSR~J1718$-$3718 could have an X-ray
luminosity as high as $5\times 10^{33}\mbox{ ergs s$^{-1}$}$, which would
give $\Lx/\Edot\approx 3$, an order-of-magnitude larger than that calculated
from nominal values \citep{zhuetal11}.
Similarly, PSR~J0726$-$2612 has $\Lx=1.5\times 10^{32}\mbox{ ergs s$^{-1}$}$
at a (uncertain) distance of 1~kpc \citep{speagleetal11};
a distance of 1.4~kpc would give $\Lx/\Edot>1$. \label{foot:etaone}}.

\begin{figure*}
\begin{center}
\includegraphics[scale=0.5]{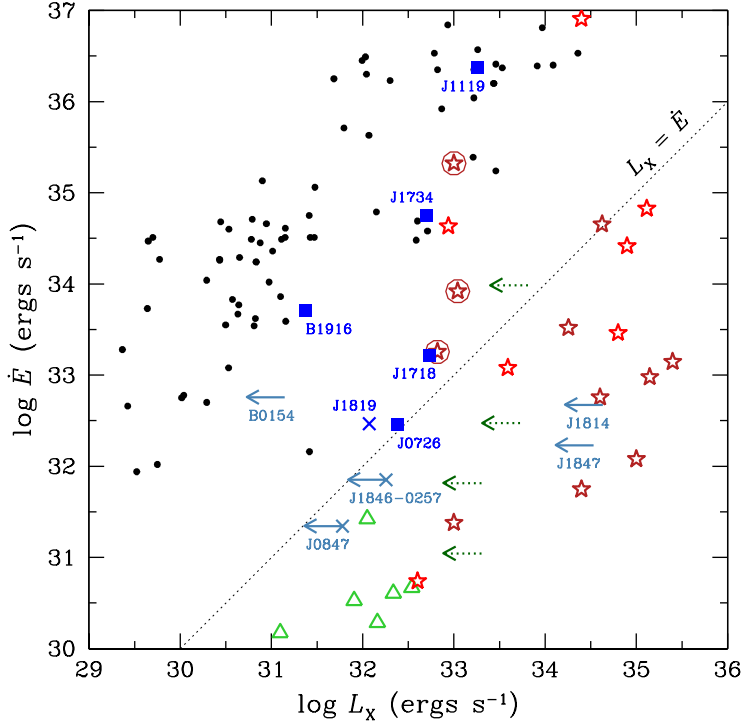}
\caption{
Rotational energy loss $\Edot$ as a function of X-ray luminosity $\Lx$.
(Red) stars denote magnetars (AXPs/SGRs), and
large circles around stars indicate magnetars that are detected in radio.
(Blue) squares and labeled upper limits indicate high-$B$ pulsars,
while crosses indicate high-$B$ pulsars which are rotating radio transients.
(Green) triangles denote ROSAT isolated neutron stars,
and dotted upper limits are for emission from the entire stellar surface
of central compact objects.
(Black) small circles denote rotation-powered radio pulsars.
Dotted line indicates $\Lx=\Edot$.
}
\label{fig:edotlx}
\end{center}
\end{figure*}

\begin{figure*}
\begin{center}
\includegraphics[scale=0.5]{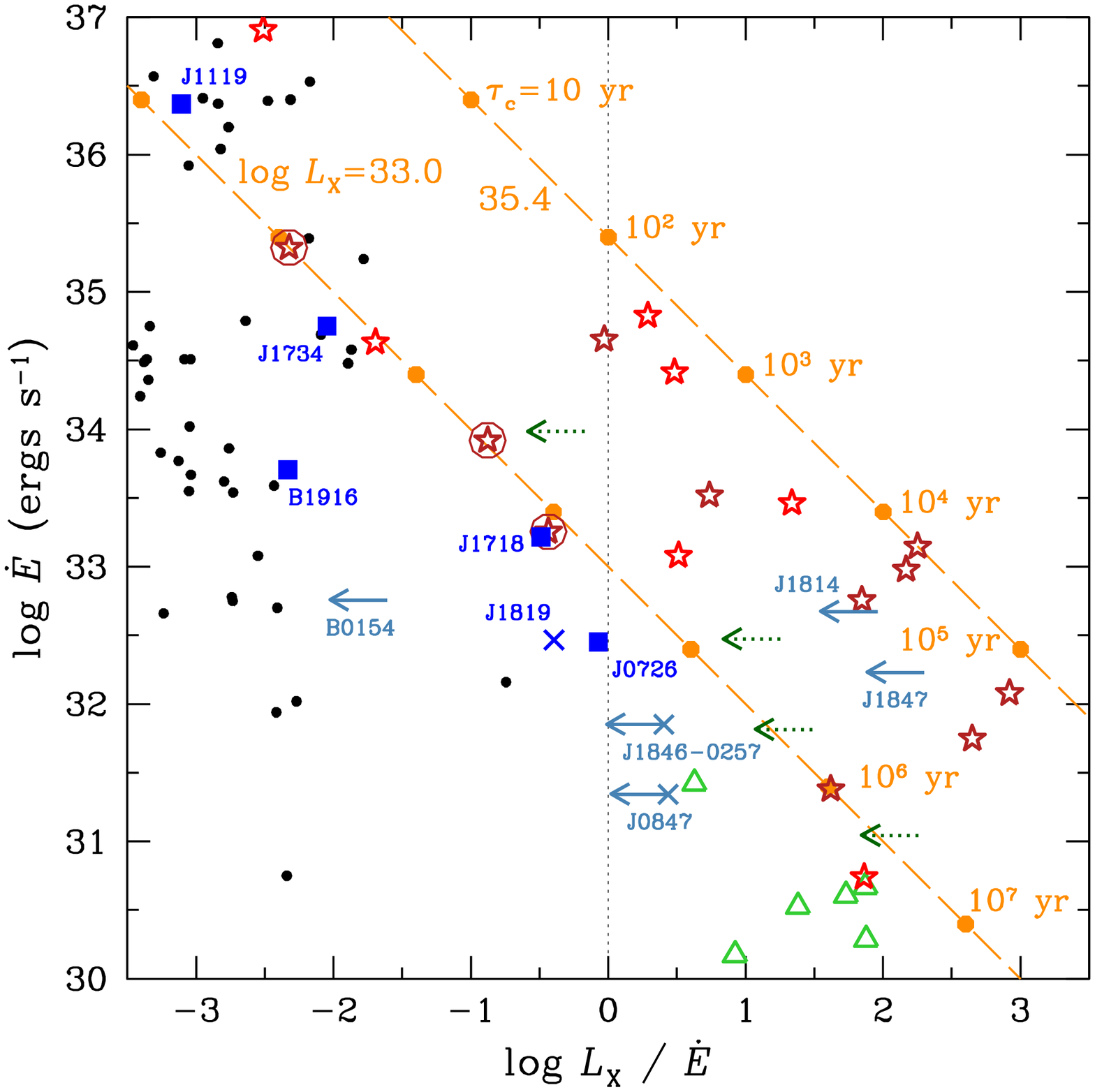}
\caption{
Rotational energy loss $\Edot$ as a function of X-ray efficiency $\Lx/\Edot$.
(Red) stars denote magnetars (AXPs/SGRs), and
large circles around stars indicate magnetars that are detected in radio.
(Blue) squares and labeled upper limits indicate high-$B$ pulsars,
while crosses indicate high-$B$ pulsars which are rotating radio transients.
(Green) triangles denote ROSAT isolated neutron stars,
and dotted upper limits are for emission from the entire stellar surface
of central compact objects.
(Black) small circles denote rotation-powered radio pulsars.
Dotted line indicates $\Lx/\Edot=1$.
Dashed lines indicate $\Lx/\Edot$ for $\log\Lx=33.0$ and $35.4$, and filled
circles give the characteristic age $\tauc$ (assuming $P=5\mbox{ s}$).
}
\label{fig:edoteta}
\end{center}
\end{figure*}

Figure~\ref{fig:edotlx} shows $\Edot$ as a function of $\Lx$ for our sources
of interest.  Also indicated is the line $\Lx=\Edot$.
It is evident that most magnetars possess X-ray luminosities which are not
powered by their rotation since $\Lx>\Edot$; this is a well-known and
previously defining property for magnetars.
Figure~\ref{fig:edoteta} shows $\Edot$ as a function of $\Lx/\Edot$, which
can be interpreted as the efficiency of converting rotational energy loss
into X-ray emission.
\citet{reaetal12} refer to this figure as the ``fundamental plane for radio
magnetars,'' although they use the electric potential difference $\Delta\Phi$
for the $y$-axis (see footnote~\ref{foot:volt}).
Our Fig.~\ref{fig:edoteta} is equivalent to the fundamental plane
since $\log\Edot=2\log\Delta\Phi+\mbox{constant}$ [see eq.~(\ref{eq:potent})].
However, by using $\Edot$, we see more clearly that the fundamental plane is
merely a rotation of Fig.~\ref{fig:edotlx}, obtained by dividing the $x$-axis
(i.e., $\Lx$) by the $y$-axis (i.e., $\Edot$); the most obvious evidence
for this rotation is the line $\Lx=\Edot$.
Thus the plane (Fig.~\ref{fig:edoteta}) containing $\Edot$ (or $\Delta\Phi$)
and $\Lx/\Edot$ is no more fundamental, and contains no additional
information, than what is already present in Fig.~\ref{fig:edotlx}.
Nothing can be said about the actual pulsar voltage gap, except as
inferred from rotational energy loss.

\citet{reaetal12} draw attention to the fact that magnetars and high-$B$
pulsars appear to lie in a relatively narrow band along the diagonal in
Fig.~\ref{fig:edoteta},
i.e., there is an anti-correlation between $\Edot$ and $\Lx/\Edot$ for these
sources.
They perform magneto-thermal simulations (see \citealt{ponsetal09}, for details)
and find evolutionary tracks that cluster approximately along this diagonal
when the stars are born with fields $>5\times 10^{13}\mbox{ G}$.
Furthermore, they find that these neutron stars evolve quickly to $\Lx/\Edot>1$,
and this rapid movement through parameter space explains the absence of
magnetars in the upper left of Fig.~\ref{fig:edoteta}.
First, we demonstrate that the clustering along the diagonal is an illusion
and due to the particular choice of axes.
In Fig.~\ref{fig:edoteta}, we plot two lines that approximately span the
observed magnetar clustering.
These two lines are simply $\Lx/\Edot$ as a function of $\Edot$ for constant
values of $\Lx$ ($=10^{33}$ and $2.5\times 10^{35}\mbox{ ergs s$^{-1}$}$);
they arise because we are plotting $\log\Edot=-\log\Edot+\mbox{constant}$.
Thus, while the magnetic field is likely to be important in determining the
X-ray brightness of a pulsar (see Sec.~\ref{sec:intro} and \ref{sec:discuss}),
one does not need detailed knowledge of the field to understand magnetars
in the parameter space of $\Edot$ and $\Lx$.

\begin{figure*}
\begin{center}
\includegraphics[scale=0.5]{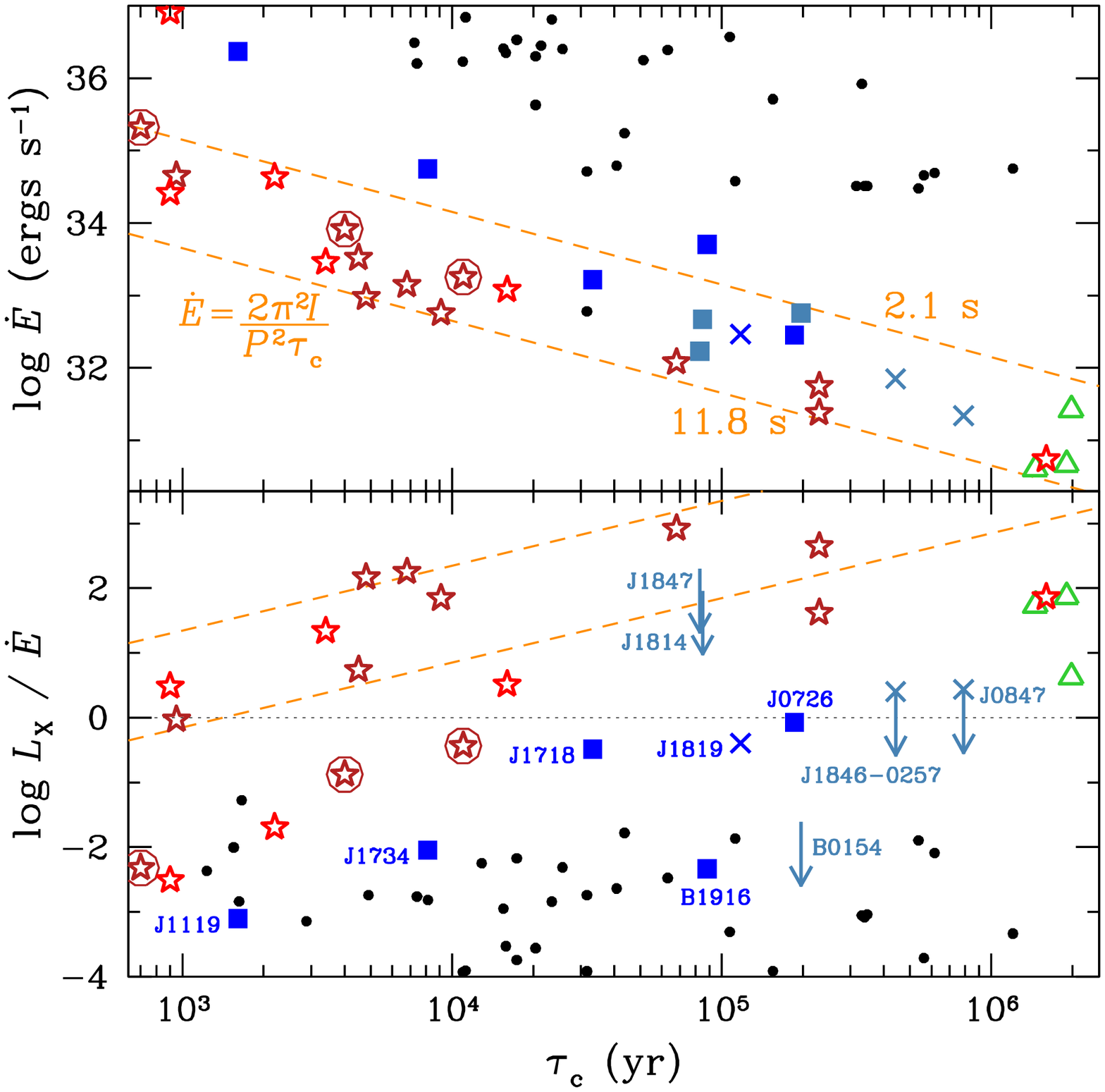}
\caption{
Rotational energy loss $\Edot$ (top panel)
and X-ray efficiency $\Lx/\Edot$ (bottom panel)
as a function of characteristic age $\tauc$.
(Red) stars denote magnetars (AXPs/SGRs), and
large circles around stars indicate magnetars that are detected in radio.
(Blue) squares and labeled upper limits indicate high-$B$ pulsars,
while crosses indicate high-$B$ pulsars which are rotating radio transients.
(Green) triangles denote ROSAT isolated neutron stars,
and dotted upper limits are for emission from the entire stellar surface
of central compact objects.
(Black) small circles denote rotation-powered radio pulsars.
Dotted line indicates $\Lx/\Edot=1$.
Dashed lines indicate $\Edot$ and $\Lx/\Edot$ (taking
$\Lx=10^{35}\mbox{ ergs s$^{-1}$}$) for spin periods $P=2.5\mbox{ s}$ and
$11.8\mbox{ s}$, where $\Edot\propto 1/(P^2\tauc)$ [see eq.~(\ref{eq:edot})].
}
\label{fig:tauc}
\end{center}
\end{figure*}

We next consider the fact that, for the neutron stars of interest here, there
are only three observables: spin period $P$, period derivative $\Pdot$, and
X-ray luminosity $\Lx$, with the first two yielding $\Edot$
[see eq.~(\ref{eq:edot})].
Thus far, we have only discussed two parameters, $\Edot$ and $\Lx$.
Along each dashed line of $\Lx/\Edot$ as a function of $\Edot$ in
Fig.~\ref{fig:edoteta}, we indicate particular values of a third parameter,
the characteristic age $\tauc$ [see eq.~(\ref{eq:tauc})],
for a typical magnetar spin period $P=5\mbox{ s}$
(magnetars have spin periods between 2.1 and 11.8~s).
It is clear that we can obtain ``evolutionary sequences,'' although no
evolution is actually taking place since $P$ is constant while $\Pdot$ is
changing along each sequence.
The lack of magnetars in the upper left of Fig.~\ref{fig:edoteta}
is the same dearth of sources seen at short characteristic ages in the
traditional $P$ versus $\Pdot$ diagram for pulsars.
In Figure~\ref{fig:tauc}, we show plots of the three observables $P$, $\Pdot$,
and $\Lx$, with the first two in the form of $\Edot$ and $\tauc$.
The point here is to illustrate that the top panel ($\Edot$-$\tauc$) encodes
no additional information than what is contained in the $P$-$\Pdot$ diagram;
instead of inhabiting the upper right in $P$-$\Pdot$ parameter space,
magnetars are in the lower left in $\Edot$-$\tauc$ because of their
longer spin periods [recall $\Edot\propto P^{-3}$; see eq.~(\ref{eq:edot})].
On the other hand, the bottom panel ($\Lx/\Edot$-$\tauc$) has additional
information because it uses an independent observable, i.e., magnetars tend
to be more X-ray luminous than normal pulsars and have $\Lx>\Edot$
(see Sec.~\ref{sec:intro}).

Let us now examine the radio {\it inactivity} criterion $\Lx/\Edot>1$
(i.e., a source which satisfies this condition will not emit in radio)
in the unification picture of neutron stars (i.e., that the very different
observed properties of rotation-powered pulsars, magnetars, and
other neutron stars are the result of a few intrinsic parameters,
e.g., age and magnetic field; \citealt{kaspi10}).
Note that \citet{reaetal12} only claim this criterion to be true for magnetars.
From Figs.~\ref{fig:edotlx}-\ref{fig:tauc},
we see that all pulsars, not just magnetars, with a measured $\Lx>\Edot$
have not been observed to emit at radio wavelengths
(for ROSAT isolated neutron stars, see \citealt{kondratievetal09}; for central
compact objects, see, e.g, \citealt{deluca08,halperngotthelf10a}, for review).
However, there are several high-$B$ radio pulsars which have X-ray luminosity
limits that still could allow them to violate the radio inactivity condition,
especially PSR~J1847$-$0130 with $\Lx/\Edot<200$ and
PSR~J1814$-$1744 with $\Lx/\Edot<90$.
The criterion $\Lx/\Edot<1$ for radio {\it activity} also appears to be
valid for most pulsars, not just magnetars;
note that \citet{reaetal12} argue that the two magnetars with $\Lx<\Edot$ and
are not seen in radio are, in actuality, radio emitters but have factors which
have thus far prevented their detection in radio.
We caution though that we use nominal values of $\Lx$ obtained from the
literature and that some of these are subject to large uncertainties due
to, e.g., unknown source distance.
In particular, the uncertainties in inferred X-ray luminosities for the
two radio pulsars PSR~J1718$-$3718 and PSR~J0726$-$2612 allow each to have
$\Lx>\Edot$ (see footnote~\ref{foot:etaone}).
Also $\Lx$ is the X-ray luminosity as measured by a distant observer.
If X-ray luminosity is a primary determinant in pulsar radio
activity/inactivity,
it is possible that the non-redshifted luminosity (or temperature) at the
neutron star surface should be the proper value to use.
Finally, it is of course possible that some sources have radio beams that
never cross our line-of-sight and thus have not been detected in radio.

\section{Discussion} \label{sec:discuss}

We showed that using the two prime pulsar observables, spin period $P$
and period derivative $\Pdot$, to calculate the voltage gap $\Delta\Phi$,
as derived from standard pulsar theory \citep{goldreichjulian69}, yields
no new information beyond what can be inferred from the spin-down
luminosity $\Edot$.
As a consequence, the parameter space or plane spanning $\Delta\Phi$ and
$\Lx/\Edot$ is a simple transformation of, and no more fundamental or
optimal than, the plane spanning $\Edot$ and $\Lx$.
We showed that trends (in $\Delta\Phi$ and $\Lx/\Edot$ parameter space)
seen among sources can be easily understood from standard pulsar
theory and do not require complex simulations for explanation.
For example, the anti-correlation between $\Delta\Phi$ and $\Lx/\Edot$
is a deception due to their particular dependence on $P$ and $\Pdot$.
Finally, we showed that a condition for pulsar radio activity/inactivity
based on X-ray efficiency ($\Lx/\Edot\sim 1$) seems to hold true in the
unified picture of neutron stars, although there exists sufficient
uncertainties for some sources that they could invalidate this conclusion.
Thus the mechanism for pulsar radio emission appears to be similar among
the different classes of neutron stars.
However there are important differences between the observed properties
of radio emission from magnetars and normal pulsars
(see \citealt{reaetal12}, and references therein),
and these are likely due in part to differences in emission location
and magnetic field strength/geometry of the magnetosphere
(see, e.g., \citealt{beloborodovthompson07,beloborodov12},
and references therein).
New sources and improved measurements would provide a better understanding
of radio behavior, as well as continuing theoretical work.

Up to this point, we have not discussed one other parameter, magnetic field
$B$, that is often used to compare magnetars and pulsars.
The most common method of determining $B$ for individual sources is by
using eq.~(\ref{eq:magb})
[or eq.~(\ref{eq:magbspit});
see also \citealt{glampedakisandersson11}, where it is argued that using
eq.~(\ref{eq:magb}) for magnetars leads to an overestimate of $B$].
In this case, our statements regarding comparisons between
parameters that are only derived from the two observables $P$ and $\Pdot$
apply here as well.  For example, plots of $B$ versus $\tauc$ provide no
additional information than what is contained within the standard pulsar
$P$-$\Pdot$ diagram,
while plots of surface temperature $T$ versus $B$
(see, e.g., \citealt{ponsetal07,zhuetal09})
and $T$ versus $\tauc$ (see, e.g., \citealt{kaplanvankerkwijk11})
are similar to $\Lx$-$\Edot$ (though there are systematic differences
between measurements of $T$ and $\Lx$).  On the other hand,
independent measurements of magnetic field (from, e.g., spectral lines)
or true ages (from, e.g., supernova remnants) do yield new information
and are thus extremely valuable.
In regards to the latter, we note that, for young pulsars, characteristic
age generally disagrees with true age in cases where both can be determined
(see, e.g., \citealt{hoandersson12}).

Finally, the reason why magnetars exhibit high X-ray luminosities $\Lx$
(for their age) is not known for certain.
What is known is that an additional source of internal heat (beyond
residual heat from neutron star formation) must be present in the outer crust
\citep{kaminkeretal06,kaminkeretal09,hoetal12}.
Magnetic field evolution and decay could provide this heat source
(see, e.g., \citealt{ponsetal07,ponsetal09,cooperkaplan10,priceetal12}).
We note that several magnetars may have very similar X-ray luminosities
(see Fig.~\ref{fig:edotlx}; see also \citealt{durantvankerkwijk06});
this could suggest that their crustal field strengths are similar and the
field decay timescale is longer than the age of the oldest of these sources.

\section*{acknowledgments}
WCGH thanks Nils Andersson and the anonymous referee for comments
and Jos\'{e} Pons and Nanda Rea for discussion and clarifications.
WCGH acknowledges support from the Science and
Technology Facilities Council (STFC) in the UK.

\bibliographystyle{mnras}

\label{lastpage}

\end{document}